# Wide Angle Dynamically Tunable Enhanced Infrared Absorption on Large Area Nanopatterned Graphene

*Alireza Safaei[1,2ǂ], Sayan Chandra[2ǂ], Michael N. Leuenberger[1,2,3], Debashis Chanda[1,2,3]**

[1]Department of Physics, University of Central Florida, Orlando, Florida 32816, USA.

[2]NanoScience Technology Center, University of Central Florida, Orlando, Florida 32816, USA.

[3]CREOL, The College of Optics and Photonics, University of Central Florida, Orlando, Florida 32816, USA.

ǂ These authors contributed equally.

*Correspondence and requests for materials should be addressed to D.C. (email: Debashis.Chanda@ucf.edu).

**Enhancing light-matter interaction by exciting Dirac plasmons on nanopatterned monolayer graphene is an efficient route to achieve high infrared absorption. Here, we designed and fabricated the hexagonal planar arrays of nanohole and nanodisk with and without optical cavity to excite Dirac plasmons on the patterned graphene and investigated the role of plasmon lifetime, extinction cross-section, incident light polarization, angle of incident of light and pattern dimensions on the light absorption spectra. By incorporating a high-k $Al_2O_3$ layer as gate dielectric for dynamic electrostatic tuning of the Fermi level, we demonstrate a record peak absorption of 60% and 90% for the nanohole and nanodisk patterns, respectively, in the atmospheric transparent infrared imaging 8 – 12 μm band with high spectral tunability. Finally, we theoretically and experimentally demonstrate, for the first time, angular dependence of both s- and p- polarized light absorption in monolayer graphene. Our results showcase the practical usability of low carrier mobility CVD-grown graphene for wide angle infrared absorption, paving the path for next generation optoelectronic devices such as photodetectors, optical switches, modulators etc.**

Graphene is one of the widely studied two dimensional materials due to its special electrical and optical properties. Various promising strategies and ideas are being proposed for optical, electrical and mechanical devices based on monolayer graphene by taking advantage of unique properties, such as high carrier mobility[1-6], fast carrier relaxation time and electrostatic tunability in the devices such as transistors[7], photodetectors[8, 9], optical switches[10], nanolasers[11] and chemical sensors[12-14]. Compared to other two-dimensional materials, large scale monolayer graphene can be grown easily using chemical vapor deposition (CVD) technique which makes real-world graphene based optoelectronic devices viable even with lower carrier mobility compare to the mechanically exfoliated flakes. However, a major bottleneck is the low light-matter interaction in graphene that needs to be enhanced. Graphene is an ultrathin semi-metal with a Dirac point in the band-structure where the conduction and valence bands cross, leading to a constant light absorption (~2.3%) in the visible regime[15] and low absorption (< 3%) in the mid-IR wavelength ranges[16]. Different strategies have been pursued to increase the interaction of the incident light with monolayer graphene, while preserving its inherent properties such as high carrier mobility and fast relaxation time. To establish the feasibility of graphene based infrared absorbers and detectors, three critical aspects need to be considered, (i) dependence of absorption to the angle of incidence, (ii) spectral tunability and selectivity for wide band operation, and (iii) polarization dependence. Ideally, polarization and angle independent absorption are desirable properties of an absorber. Coupling the near-field of a metallic metasurface or photonic crystal to graphene[10, 17-20] is an indirect solution to increase the light-graphene interaction. However, in this scheme majority of incident light is dissipated as resistive ohmic loss in the metal, defeating the purpose. Additionally, owing to asymmetric metasurface designs most of the absorbers reported till date are sensitive to the polarization of incident light and exhibit limited spectral tunability[21-24]. In another approach, coupling a pristine[25] and nanopatterned[26-28] monolayer graphene to an optical cavity has been implemented to enhance absorption. Although, these recent works have experimentally demonstrated the enhancement of absorption for normal incident light[21-23, 28, 29], little is known about their absorption at higher angles of incidence, incident polarization or pattern edge states in combination with doping level.

In the infrared domain, exciting Dirac plasmons[30] on nanopatterned graphene has been adopted as a route to couple and concentrate the incident light directly on the surface thereby enhancing the infrared absorption[27,29, 31]. Depending on the nanopattern design, the Dirac plasmons on graphene can be propagating surface plasmon polaritons (SPP) or localized surface plasmons (LSPs) modes. Plasmons are qualitatively characterized by their lifetimes. Longer lifetime results in stronger electric field confinement, which manifests as higher and sharper absorption (lower FWHM) in the spectral response. It has been reported that nanopatterning of graphene introduces graphene edges, which play a vital role in modifying the light absorption spectrum. Edge scattering effects, radiative and non-radiative decay arising from Landau damping through interband and intraband transitions contribute collectively to increase the decay rates of the plasmonic excitations[28, 30, 32-35]. Therefore, the fundamental question that arises is how critical is the role of graphene edge on the localized surface plasmon (LSP) excitation, decay rate and overall absorption behavior of patterned graphene? Here, in order to investigate these aspects we identified complementary nanostructures, i.e. nanoholes and nanodisks, such that the qualitative nature of the edges are similar. A direct comparison of these complementary structures elucidates the differences in the plasmonic excitations, the degree of electrostatic spectral tunability and polarization dependences as a function of incident angle. We adopt a combinatorial investigation using theory and experiment to gain insight into the underlying physical phenomena. Based on this understanding, here we demonstrate the effect of graphene nanopatterning and the edges on the plasmon lifetime and the light absorption. The maximum achieved light absorption is experimentally measured 90% (60%) for the cavity coupled graphene nanodisk (nanohole) array for the specified geometry parameters and $E_F$= -1eV which is independent of the light polarization. An ideal strategy to enhance the light-graphene interaction should be independent of the angle of the incident light ($\theta_i$). We measured for the first time the dependence of the light absorption as a function of $\theta_i$ for s- and p-polarized light and showed they are in very good agreement with the simulated results. At the end we show the peak wavelength and moreover the magnitude of the absorption of the unpolarized light by the patterned graphene are almost independent of the angle of incidence for $\theta_i < 50^\circ$.

The architecture of the proposed graphene absorber is illustrated in Fig. 1a. Pristine graphene grown on copper foil by CVD method was transferred on $Si^{++}$ (100 μm)/$Al_2O_3$ (15 nm)/ITO (30 nm) substrate. The hole/disc diameter (D) and period (P) were varied to tune the LSPR at a desired wavelength whereby the cavity length was chosen to satisfy the quarter wavelength condition $L= m\lambda/4n_{eff}$. At this condition, constructive interference of the incident and reflected electric fields on graphene sheet intensifies the LSPs and enhances light-matter interaction, as shown in Fig. S1 and in our earlier work[28]. Here, $L$ is the cavity thickness, $n_{eff}$ is the effective refractive index of the optical cavity spacer, $m$ is the m-th order of the cavity mode and $\lambda$ is the light wavelength. The optical response of the designed graphene absorber was simulated by finite-difference time-domain (FDTD) approach. In these simulations, the mobility of graphene was chosen to be a modest 500 $cm^2/V.s$ (scattering rate $\Gamma$= 0.02 eV) to closely resemble the experimentally measured mobility (see SI), and the optical constants of graphene corresponding to different Fermi energies were calculated using the Drude model[2, 28]. The total light absorption spectra of the optical cavity-coupled nanopatterned graphene are different for the complementary graphene nanodisk and nanohole arrays with P = 400 nm, D = 200 nm and *L* = 1.3 µm, as shown in Fig. 1b, such that the peak bandwidth of the graphene nanodisk is smaller than that of its complementary graphene nanohole array. Interestingly, while the plasmon lifetime, which is inversely proportional to the FWHM of the absorption peak in nanodisks, is longer, it does not translate to a higher absorption. At resonance the extinction cross-section of any nanopattern exceeds the geometrical area by several factors[27], which scales differently for the complementary nanohole and nanodisk patterns. This in turn determines the effective absorption amplitude, which is higher in case of the nanohole pattern, in spite of shorter plasmon lifetimes. Therefore, the presence of surface plasmons breaks the symmetry of the complementary nanohole and nanodisk arrays due to their different plasmon decay rates, which suggests that maximum light absorption for respective patterns can be achieved by optimizing the geometrical area of the nanopattern such that the extinction cross-section is highest.

To obtain the maximum light absorption for the cavity-coupled graphene nanohole and nanodisk arrays, the optical response for different geometrical parameters (period and diameter) were simulated for a fixed Fermi energy ($E_F$ = -1 eV) at normal angle of incidence. For the nanohole array, a reduction in edge-to-edge distance can be achieved by either increasing the diameter for a fixed period or decreasing the period for a given diameter. By applying both strategies, we observed that the reduction in the edge-to-edge distance leads to a blue shift in the LSPR frequency, as predicted by Eq. 1 and shown in Fig. 2a-b. In this case, the device parameters were optimized to obtain maximum absorption at $\lambda_{res}$ = 8µm.

The LSPR frequency of the nanopatterned graphene is given by[28, 29, 31]

$$\omega_{res} = \mathcal{B}\sqrt{\frac{\Im E_F}{d}}, \tag{1}$$

where $\mathcal{B}$ is a constant value, $d$ is the edge-to-edge distance of the nanopattern and $\Im$ is the eigenvalue of the self-consistent total electric potential equation. For the graphene nanodisk pattern, the increase in diameter for a constant period results in a red shift of the LSPR wavelength, as shown in Fig. 2c-d and predicted by Eq. (1), while the decrease in the period for a constant diameter enhances the far-field and near-field coupling of the nanodisks, giving rise to a slight red shift. The amount of light absorption on the graphene nanodisk array depends on the density ratio ($\rho_r$), which is defined as the ratio of the graphene area to the unit cell area. The density ratio can be enhanced by increasing the diameter or decreasing the period. For a cavity thickness of $L$= 1.5 μm, the light absorption was found to be a continuous increasing function of the diameter and a decreasing function of period, as shown from Fig. 2c-d.

To validate the simulated results, absorber devices were fabricated using period and diameter values that yielded highest absorption for nanoholes and nanodisks respectively. The nanohole and nanodisk arrays are patterned on the transferred monolayer graphene by electron beam lithography (EBL) followed by oxygen RIE. The scanning electron microscopy (SEM) images in Fig. 3a shows the fabricated hole and disk arrays in the transferred graphene. A layer of semi-transparent SU-8 photoresist polymer was spun-coated on the patterned graphene to form the optical cavity followed by 2 hours UV-exposure and one-hour baking (95 °C). A hard layer of $Al_2O_3$ (50 nm) is deposited on SU-8 to protect SU-8 from metal deposition and an optically thick layer of gold (200 nm) was electron beam evaporated to form the back mirror. The $Si^{++}$ (100 μm) sheet used as the back gate has ~ 70% transmission in mid-IR range, as shown in Fig. S3 and light is incident from the silicon side. During the transfer and fabrication process, the monolayer graphene was found to be chemically self-doped as p-type with $E_F$~ -0.6 eV that can be attributed to residual polymer and presence of p-type $Al_2O_3$ dielectric layer[36, 37]. Electrostatic tunability of $E_F$ was achieved by applying a voltage across the $Al_2O_3$ layer such that a negative voltage resulted in accumulation of positive charges (holes) thereby driving the system to a higher Fermi level. The Fermi energy of the nanopatterned graphene can be altered from -0.55 eV to -1.0 eV as shown in Fig. 3b which is associated with a decrease in channel resistance by a factor of 3. Increase in the Fermi energy of graphene to $|E_F|$= 1.0 eV is necessary to intensify the electric dipoles oscillations that allow more photons to couple to the patterned graphene edges. Our choice of high-k hard dielectric ($Al_2O_3$) for electrostatically doping graphene provides superior chemical stability in time over the more commonly used ion-gel (see Fig S8a,b).

Raman spectroscopy was performed to confirm the quality of graphene before and after nanopatterning. As shown in Fig 3c, the characteristic 2D and G bands of graphene are visible for pristine and patterned graphene, however a red shift ($\Delta\omega \sim 7\ cm^{-1}$) in the spectrum for patterned graphene was observed which suggests a modification in the dispersion function of the acoustic and optical phonons due to the nanopatterning[38, 39] (see SI).

The normal-angle reflection spectra of the fabricated cavity-coupled graphene nanohole and nanodisk absorbers were measured using a Bruker Vertex 80 Fourier transform infrared spectrometer (FTIR). The light reflection from the absorber stack without patterned graphene, *i.e.* $Si^{++}$ (100 µm)/$Al_2O_3$ (15 nm)/ITO (30 nm)/SU-8 ($L$)/$Al_2O_3$ (50 nm)/gold (200 nm) was taken as the reference and the light absorption spectra was calculated as A= 1-R. As shown in Fig. 4a, the light absorption of the graphene nanohole array reaches ~ 60% (at $E_F$ = -1 eV), which is 35% higher than the previously reported maximum absorption in the 8 – 12 µm band[28]. Electrostatic tunability of ~2.46 µm is observed by changing $E_F$ from -0.55 eV to -1 eV. A near perfect absorption of 90% was recorded for the nanodisk array, which is electrostatically tunable over a spectral width of ~1.11 μm. There is in very good agreement with the simulated results. Increase in the Fermi energy to the negative values means more hole density and creation of stronger electric dipoles on the patterned graphene which results in enhanced light absorption along with a blue shift in the LSPR frequency as depicted in Fig. 4c. The extraordinary near-field enhancement by factors of 500 and 1100 (inset of Fig. 4 a and b) for nanohole and nanodisk arrays, respectively, explain the unprecedented high light-matter interaction and infrared absorption values recorded experimentally. The measured light absorption spectra of the patterned graphene without optical cavity is shown (Fig S4) along with simulated spectra (A=1-T-R) to further elucidate the excellent agreement between experiment and theory.

To validate the operability of any absorber, it is critical to investigate its angular dependence to light. In the seminal work by Thongrattanasiri *et. al.*, it was analytically shown for periodically patterned graphene that under the condition of no transmission, the angular optical response to light, which depends on the polarization, is primarily determined by its mobility and extinction cross-section. Enhanced absorption would necessitate maximizing the extinction cross-section, which can be achieved by pattern optimization such that the decay rate ($\kappa$) is much higher than the radiative ($\kappa_r$) contribution ($\kappa \gg \kappa_r$). The decay rates along with the plasmon frequency $\omega_p$ determines the graphene polarizability given by

$$\alpha(\omega) = \frac{3c^3\kappa_r}{2\omega_p^2}\frac{1}{\omega_p^2-\omega^2-i\kappa\omega^3/\omega_p^2}\,. \qquad (2)$$

Under the assumption that the polarizabilities of monolayer graphene nanohole/nanodisk are almost independent of the angle of incidence[29, 40], the LSPR frequency is expected to not be affected by change in incident angle of light. This is confirmed by the FDTD simulations where the LSPR frequency is found to be almost independent of $\theta_{\text{inc}}$, for both nanohole and nanodisk array patterned graphene with and without optical cavity (See Fig S6 and S7). At higher angles of incidence, the peak absorption of the system differs for the in-plane polarization (p-polarization) and the out-of-plane (s-polarization) light. For p-polarized light, the absorption magnitude scales with the electric field component parallel to the surface, $\boldsymbol{E}_i cos\theta_i$ as the angle of incidence increases. In case of s-polarized light, the LSPR frequency is almost independent of the angle of incidence for the nanohole and nanodisk array devices in the cavity coupled (Fig 5) and no-cavity (Fig S6 and S7) systems. However, the magnitude of absorption increases with angle of incidence for s-polarized light in contrast to the results obtained for p-polarized light shown in figures S6 and S7. While the magnitude of electric field parallel to graphene surface remain unaffected for all angles of incidence (see inset of Fig 5c), the scattering cross-section increases and scales as $\boldsymbol{E}_i sin\theta_i$. This explains the enhancement in absorption as the angle of incidence increases (Fig S6 and S7). Unlike the no-cavity nanodisk array absorber, the cavity couple system exhibits an increase in absorption for $0^{\circ} < \theta_i < 50^{\circ}$ but for higher angles, the absorption drops (Figure 5). The temporal and spatial interference between the optical cavity and the graphene plasmonic modes modifies the angular response such that for incident angles $\theta_i > 50^{\circ}$, destructive interference of the incident ($\mathbf{E}_i$) and reflected electric fields ($\mathbf{E}_r$) arising from phase difference lowers the light absorption, as shown in Fig. 5c. Such behavior in the angular response of cavity-coupled absorbers for s-polarized light is not uncommon and was previously shown in a $VO_2$ based system[41]. The FDTD predictions are well supported by experimental data (shown in Figure 5, S6 and S7) acquired using an integrating sphere coupled to a FTIR.

While the calculations by Thongrattanasiri *et. al.*, were done for ideal graphene with mobility of 10 000 cm /Vs, we show that it can be extended to CVD grown low mobility graphene and successfully models the FDTD and experimental results discussed below. Using Eq. (2) to fit the FDTD results at normal incidence, we obtain $\hbar\kappa = 3.5 \times 10^{-2} eV/6.9 \times 10^{-2} eV$ and $\hbar\kappa_r = 1.22 \times 10^{-4} eV/4.5 \times 10^{-4} eV$ for the graphene nanodisk/nanohole array respectively (see Fig. S10) which satisfies the criteria, $\kappa \gg \kappa_r$. Following that, the absorption of the patterned graphene for various incident angles can be calculated via the total light reflection coefficient of the cavity-coupled patterned graphene ($A = 1 - |\mathcal{R}|^2$)

$$\mathcal{R} = r_0 + \frac{r(1 \pm r_0)^2}{(1 - r r_0)} \ , \qquad (3)$$

where, $r_0$ is the Fresnel reflection coefficient of the cavity spacer without graphene and the reflection coefficient of the patterned graphene is given by, $r = \pm iS/(\alpha^{-1} - G)$ [27, 42] for the arrays with periods much smaller than the wavelength ($P \ll \lambda$). The lattice sum for this condition is reduced to $G = 5.52/P^3 + i(S - 2(\omega/c)^3/3)$ for hexagonal array, where $S$ is a polarization-dependent parameters, i.e. $S_s = 2\pi\omega/(cA\ cos\theta_i)$, $S_p = 2\pi\omega cos\theta_i/(cA\ )$ and $A$ is the unit-cell area[27]. The calculated peak absorption as a function of incident angle for s and p polarized light is overlaid on the FDTD and experimentally obtained results as shown in Figure 6. Clearly a good agreement between the analytical, simulation and experimental data is evident. Figure 6 shows the maximum absorption for different incident angles at the resonance wavelength shown by the green dash line in Fig. 5. The results of the unpolarized light (Fig. 6-bottom) shows that the maximum light absorption of the graphene absorber is almost independent of the incident angle for $\theta_i \leq 50°$.

In conclusion, we have investigated infrared absorption in an optical cavity-coupled low carrier mobility (μ= 500 $cm^2/V.s$) CVD-grown graphene with hexagonal array of nanoholes and nanodisks in the infrared transparent 8 -12 μm band. Due to the differences in the extinction cross-section of nanohole and nanodisks for the same diameter, the plasmonic excitation on their respective edges are different, resulting in qualitatively dissimilar absorption peak profiles. A series of numerical simulations were performed to maximize infrared absorption by scanning over parameters like period and diameter for the hexagonal array of nanohole and nanodiscs. The optimized devices exhibit a record absorption of 60% for the nanohole array and up to 90% for nanodisk array when the Fermi level of graphene is increased to -1 eV by electrostatic p-doping. Such high absorption is attributed to strong plasmonic excitations at the patterned nano-edges where the localized electric field is amplified by factors of 500 for the nanohole and 1100 for the nanodisk arrays. The Fermi level of the patterned graphene is tuned by applying a voltage across a 15 nm thick layer of $Al_2O_3$, which serves as a hard gate dielectric. The high-k $Al_2O_3$ is found to be more stable over time compared to the commonly used ionic-gel gate which tend to chemically degrade in few days. A remarkable dynamic spectral tunability of 2.46 μm for nanoholes and 1.11 μm for nanodisks is achieved. It is to be stressed that the use of industry standard $Al_2O_3$ for capacitive electrostatic gating in our devices makes it a potential candidate for integration with optical design boards unlike many previously reported device architectures that are bulky due to presence of ion-gel based components[29, 43].

Finally, by using the optimal devices with highest absorption for nanoholes and nanodisks, we show for the first time a systematic angle dependent (0˚ - 70˚) optical study in the infrared domain. Although the LSPR frequency is independent of the incident angle of light, the evolution of peak absorption for s- and p-polarized light are qualitatively dissimilar, which is attributed to the different scattering cross-sections that the electric field of incident light interacts with on patterned graphene. However, the peak absorption for unpolarized light remains within 5% of its maximum up to $\theta_i \leq 50^\circ$, which suggests that both the nanoholes and nanodisk array can be operated over a wide range of angles. These angle dependent results are the first experimental validation of the theoretical model for patterned graphene devices developed by Thongrattanasiri, et al[27] based on coupled-dipole approximation. From an application point of view, the key difference between the nanohole/nanodisk arrays is the presence/absence of electrical continuity in graphene. Therefore, while the near perfect absorption of ~ 90% in the nanodisk array can be beneficial for application like wide angle optical modulators, tunable infrared camouflage, etc., the nanohole array offers applicability in next generation wide band, wide angle photodetectors based on electron-hole pair generation by exciting electrostatically tunable plasmons.

# Methods

### Device Fabrication Process

A pristine graphene grown on a 25 μm thick copper foil by CVD method transferred on $Si^{++}$ (100 μm)/$Al_2O_3$ (15 nm)/ITO (30 nm) substrate. The hard gate-dielectric ($Al_2O_3$) is grown on $Si^{++}$ by atomic layer deposition (ALD). The ITO layer is sputtered on $Al_2O_3$ via RF AJA sputtering system. The nanohole and nanodisk arrays are patterned on the transferred monolayer graphene by using electron beam lithography (EBL) following by oxygen RIE etching and dissolving the electron resist PMMA in acetone. A layer of semi-transparent SU-8 photoresist polymer as the optical cavity slab is span-coated on the patterned graphene, following by 2 hours UV-exposing and one-hour baking (95 ˚C). A hard layer of $Al_2O_3$ (50 nm) is deposited on SU-8 to protect it against meta deposition and an optically thick layer of gold (200 nm) as the back mirror is deposited on top of that.

## Materials Characterization and Measurement

The theoretical simulations are done by finite-difference time-domain (FDTD) method using Lumerical FDTD (Lumerical Inc.) software. The Raman spectrum of the grown graphene sheet is measured by WITec Renishaw RM 1000B Micro-Raman Spectrometer with an excitation laser wavelength of 514 nm and a 50x objective lens. The real and imaginary parts of the gold dielectric function used in simulations are taken from Palik[44]. The corresponding normal and non-normal incidence optical absorption measurements are performed with a integrated sphere-coupled microscope-coupled FTIR (Bruker Inc., Hyperion 1000-Vertex 80). The gate-dependent electrical conductivity is measured by using the model 2602B Keithley dual-channel system SourceMeter instrument through source-drain using two probes and the gate voltage applied using the other probes. The scanning electron microscopy is measured with Zeiss ULTRA-55 FEG SEM.

## References

1. Bonaccorso, F.; Sun, Z.; Hasan, T.; Ferrari, A. C. *Nature Photonics* **2010,** 4, (9), 611-622.
2. Falkovsky, L. A. *J Exp Theor Phys+* **2008,** 106, (3), 575-580.
3. Falkovsky, L. A.; Pershoguba, S. S. *Physical Review B* **2007,** 76, (15).
4. Falkovsky, L. A.; Varlamov, A. A. *Eur Phys J B* **2007,** 56, (4), 281-284.
5. Singh, V.; Joung, D.; Zhai, L.; Das, S.; Khondaker, S. I.; Seal, S. *Progress in Materials Science* **2011,** 56, (8), 1178-1271.
6. Hemmatiyan, S.; Polini, M.; Abanov, A.; MacDonald, A. H.; Sinova, J. *Physical Review B* **2014,** 90, (3).
7. Schwierz, F. *Nature nanotechnology* **2010,** 5, (7), 487-96.
8. Sun, Z.; Chang, H. *ACS nano* **2014,** 8, (5), 4133-56.
9. Freitag, M.; Low, T.; Zhu, W.; Yan, H.; Xia, F.; Avouris, P. *Nature communications* **2013,** 4, 1951.
10. Yao, Y.; Shankar, R.; Kats, M. A.; Song, Y.; Kong, J.; Loncar, M.; Capasso, F. *Nano Lett* **2014,** 14, (11), 6526-32.
11. Sun, Z.; Hasan, T.; Torrisi, F.; Popa, D.; Privitera, G.; Wang, F.; Bonaccorso, F.; Basko, D. M.; Ferrari, A. C. *ACS nano* **2010,** 4, (2), 803-10.
12. Lu, Y.; Goldsmith, B. R.; Kybert, N. J.; Johnson, A. T. C. *Applied Physics Letters* **2010,** 97, (8), 083107.
13. Yavari, F.; Koratkar, N. *The journal of physical chemistry letters* **2012,** 3, (13), 1746-53.
14. Singh, E.; Meyyappan, M.; Nalwa, H. S. *ACS applied materials & interfaces* **2017,** 9, (40), 34544-34586.
15. Nair, R. R.; Blake, P.; Grigorenko, A. N.; Novoselov, K. S.; Booth, T. J.; Stauber, T.; Peres, N. M.; Geim, A. K. *Science* **2008,** 320, (5881), 1308.
16. Yan, H.; Xia, F.; Zhu, W.; Freitag, M.; Dimitrakopoulos, C.; Bol, A. A.; Tulevski, G.; Avouris, P. *ACS nano* **2011,** 5, (12), 9854-60.
17. Gan, X.; Shiue, R. J.; Gao, Y.; Mak, K. F.; Yao, X.; Li, L.; Szep, A.; Walker, D., Jr.; Hone, J.; Heinz, T. F.; Englund, D. *Nano Lett* **2013,** 13, (2), 691-6.
18. Otsuji, T.; Popov, V.; Ryzhii, V. *J Phys D Appl Phys* **2014,** 47, (9), 094006.
19. Zhang, Y.; Feng, Y.; Zhu, B.; Zhao, J.; Jiang, T. *Optics express* **2014,** 22, (19), 22743-52.
20. Majumdar, A.; Kim, J.; Vuckovic, J.; Wang, F. *Ieee J Sel Top Quant* **2014,** 20, (1).

21. Kim, S.; Jang, M. S.; Brar, V. W.; Mauser, K. W.; Kim, L.; Atwater, H. A. *Nano Lett* **2018,** 18, (2), 971-979.
22. Brar, V. W.; Jang, M. S.; Sherrott, M.; Lopez, J. J.; Atwater, H. A. *Nano Lett* **2013,** 13, (6), 2541-7.
23. Ju, L.; Geng, B.; Horng, J.; Girit, C.; Martin, M.; Hao, Z.; Bechtel, H. A.; Liang, X.; Zettl, A.; Shen, Y. R.; Wang, F. *Nature nanotechnology* **2011,** 6, (10), 630-4.
24. Safaei, A.; Vázquez-Guardado, A.; Franklin, D.; Leuenberger, M. N.; Chanda, D. *Advanced Optical Materials* **2018**, 1800216.
25. Furchi, M.; Urich, A.; Pospischil, A.; Lilley, G.; Unterrainer, K.; Detz, H.; Klang, P.; Andrews, A. M.; Schrenk, W.; Strasser, G.; Mueller, T. *Nano Lett* **2012,** 12, (6), 2773-7.
26. Alaee, R.; Farhat, M.; Rockstuhl, C.; Lederer, F. *Optics express* **2012,** 20, (27), 28017-28024.
27. Thongrattanasiri, S.; Koppens, F. H.; Garcia de Abajo, F. J. *Phys Rev Lett* **2012,** 108, (4), 047401.
28. Safaei, A.; Chandra, S.; Vázquez-Guardado, A.; Calderon, J.; Franklin, D.; Tetard, L.; Zhai, L.; Leuenberger, M. N.; Chanda, D. *Physical Review B* **2017,** 96, (16).
29. Fang, Z.; Thongrattanasiri, S.; Schlather, A.; Liu, Z.; Ma, L.; Wang, Y.; Ajayan, P. M.; Nordlander, P.; Halas, N. J.; Garcia de Abajo, F. J. *ACS nano* **2013,** 7, (3), 2388-95.
30. Paudel, H. P.; Safaei, A.; Leuenberger, M. N., Nanoplasmonics in Metallic Nanostructures and Dirac Systems. In *Nanoplasmonics - Fundamentals and Applications*, Barbillon, D. G., Ed. InTech: 2017.
31. Gao, W.; Shu, J.; Qiu, C.; Xu, Q. *ACS nano* **2012,** 6, (9), 7806-13.
32. Yan, H. G.; Low, T.; Zhu, W. J.; Wu, Y. Q.; Freitag, M.; Li, X. S.; Guinea, F.; Avouris, P.; Xia, F. N. *Nature Photonics* **2013,** 7, (5), 394-399.
33. Fei, Z.; Goldflam, M. D.; Wu, J. S.; Dai, S.; Wagner, M.; McLeod, A. S.; Liu, M. K.; Post, K. W.; Zhu, S.; Janssen, G. C.; Fogler, M. M.; Basov, D. N. *Nano Lett* **2015,** 15, (12), 8271-6.
34. Brown, A. M.; Sundararaman, R.; Narang, P.; Goddard, W. A., 3rd; Atwater, H. A. *ACS nano* **2016,** 10, (1), 957-66.
35. Sundararaman, R.; Narang, P.; Jermyn, A. S.; Goddard, W. A., 3rd; Atwater, H. A. *Nature communications* **2014,** 5, 5788.
36. Shin, S.; Kim, S.; Kim, T.; Du, H.; Kim, K. S.; Cho, S.; Seo, S. *Carbon* **2017,** 111, 215-220.
37. Deng, C.; Lin, W.; Agnus, G.; Dragoe, D.; Pierucci, D.; Ouerghi, A.; Eimer, S.; Barisic, I.; Ravelosona, D.; Chappert, C.; Zhao, W. *The Journal of Physical Chemistry C* **2014,** 118, (25), 13890-13897.
38. Maultzsch, J.; Reich, S.; Thomsen, C.; Requardt, H.; Ordejon, P. *Phys Rev Lett* **2004,** 92, (7), 075501.
39. Qian, J.; Allen, M. J.; Yang, Y.; Dutta, M.; Stroscio, M. A. *Superlattices and Microstructures* **2009,** 46, (6), 881-888.
40. Maier, S. A. **2007**.
41. Kocer, H.; Butun, S.; Palacios, E.; Liu, Z.; Tongay, S.; Fu, D.; Wang, K.; Wu, J.; Aydin, K. *Scientific reports* **2015,** 5, 13384.
42. García de Abajo, F. J. *Reviews of Modern Physics* **2007,** 79, (4), 1267-1290.
43. Cho, J. H.; Lee, J.; Xia, Y.; Kim, B.; He, Y.; Renn, M. J.; Lodge, T. P.; Frisbie, C. D. *Nature materials* **2008,** 7, (11), 900-6.
44. Palik, E. D. *J Opt Soc Am A* **1984,** 1, (12), 1297-1297.

# ASSOCIATED CONTENT

### Supporting Information

Additional information including the simulated absorption spectra versus cavity thickness for the graphene nanodisk and nanohole arrays (different geometries), the measured light transmission of the $Si^{++}$ substrate, the simulated and measured absorption spectra for the patterned graphene without optical cavity, the simulated and measured results of the absorption spectra of angled incident light by the samples with and without cavity, characterization and comparison of the hard ($Al_2O_3$) and soft (ion-gel) gate dielectrics, analysis of electrostatic doping of graphene and the graphene plasmon decay rates.

## Author information

### Corresponding Author

*Correspondence and requests for materials should be addressed to D.C. (email: Debashis.Chanda@creol.ucf.edu).

### Author Contributions

A.S. and D.C. conceived the idea. A.S. designed and performed the experiments. S.C. provided technical assistance. A.S., S.C., D.C. analyzed and simulated the data. M.L. and D.C. contributed materials/analysis tools. A.S., S.C. and D.C. co-wrote the paper.

### Notes

The authors declare no competing financial interests.

## Acknowledgments

This work at University of Central Florida was supported by DARPA under the WIRED program grant no. HR0011-16-1-0003.

## Figures

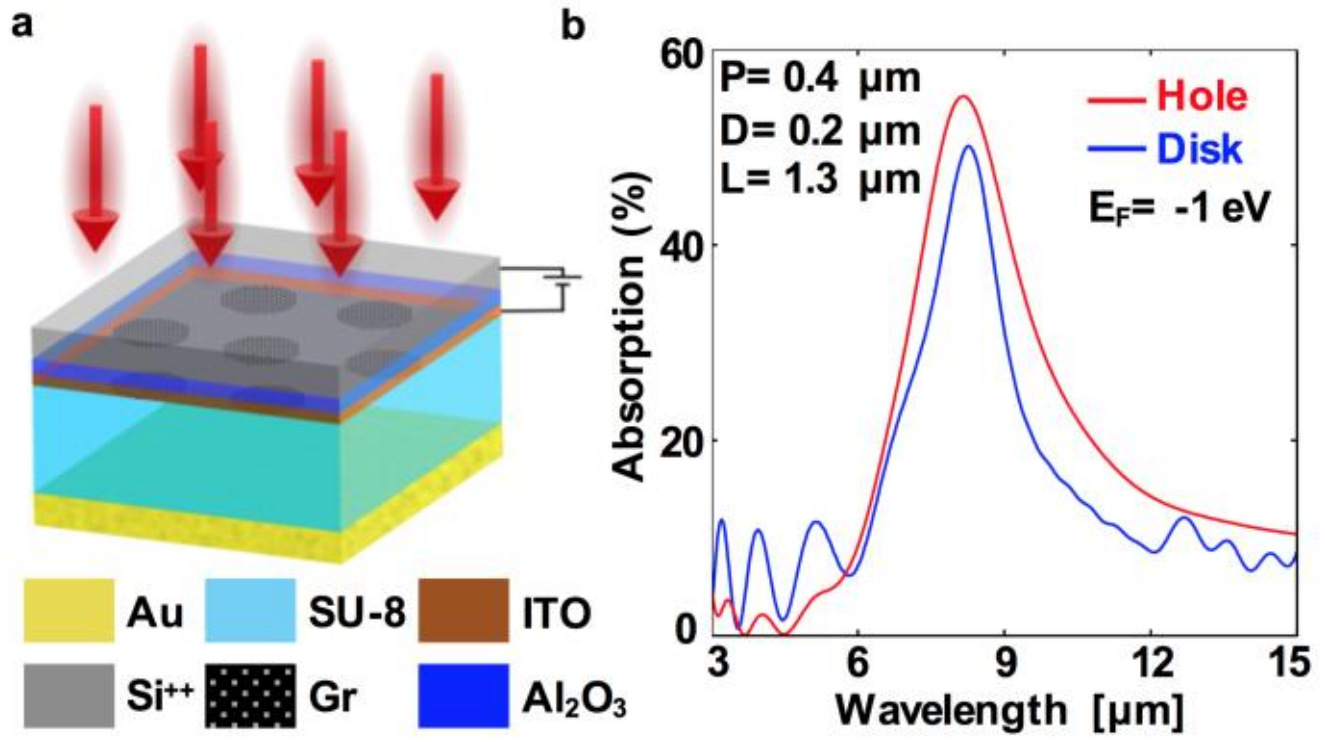

**Figure 1|** (a) Schematic of the optical cavity-coupled patterned graphene. (b) The light absorption spectra of the optical cavity-coupled graphene hole (red) and disk (blue) arrays with P= 400 nm, D= 200 nm, L= 1300 nm and $E_F$= -1 eV.

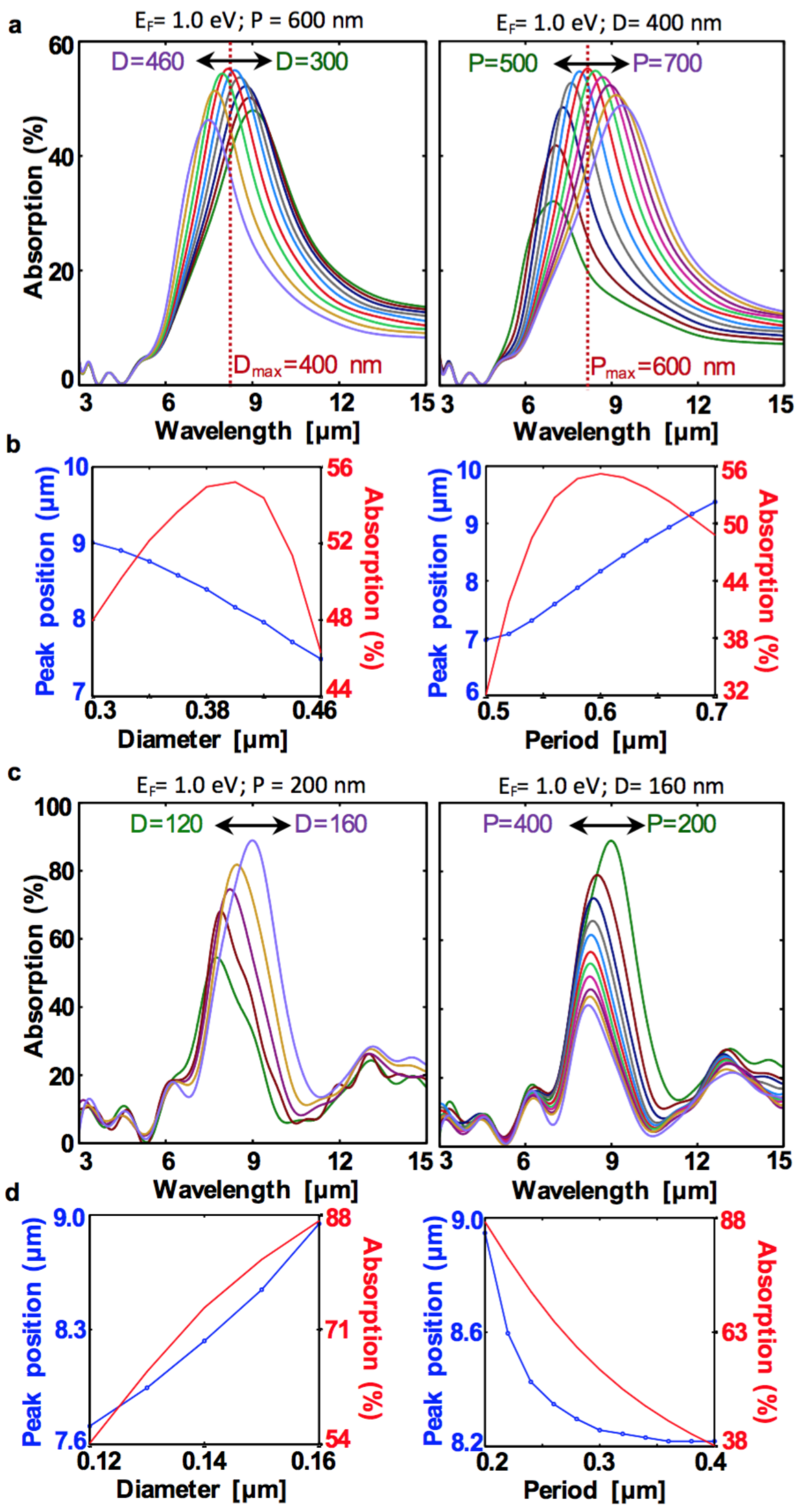

**Figure 2| Geometrical tunability.** (a) The light absorption of the cavity-coupled nanohole graphene with the thickness L= 1.3 μm for different diameters in a constant period (P= 600 nm) (left) and different periods in a constant diameter (D= 400 nm) (right). (b) The peak position and the absorption for the graphene nanohole as a function of diameter (left) and period (right). (c) The light absorption of the cavity-coupled nanodisk graphene with the thickness L= 1.5 μm for different diameters in a constant period (P= 200 nm) (left) and different periods in a constant diameter (D= 160 nm) (right). (b) The peak position and the absorption for the graphene nanodisk as a function of diameter (left) and period (right).

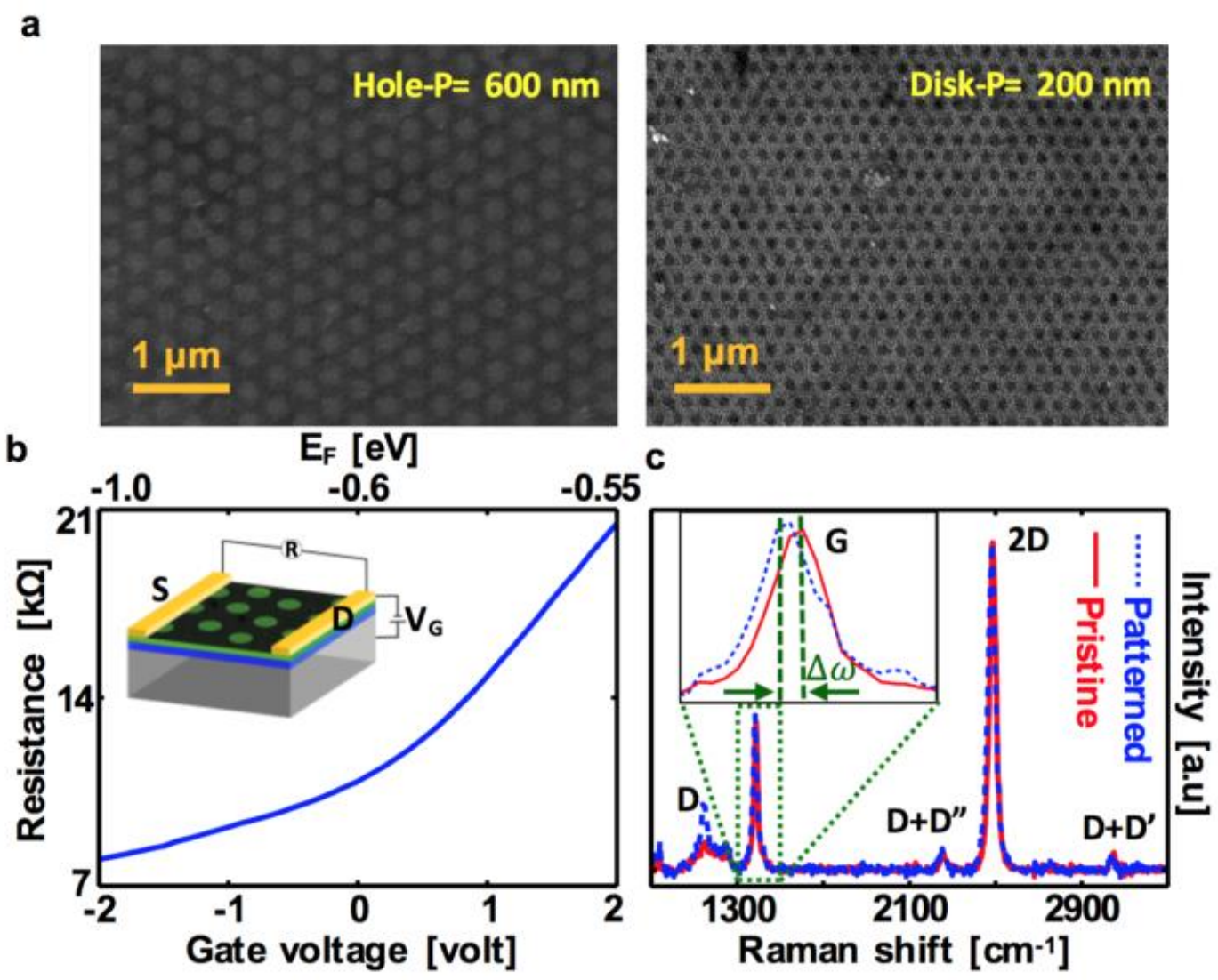

**Figure 3| Fabrication and characterization.** (a) SEM image of the fabricated graphene nanohole (left) and nanodisk (right). (b) The electrical resistance of the patterned graphene as a function of the gate voltage. (c) The Raman spectroscopy of the pristine and patterned monolayer graphene ($E_F$= -0.7 eV). The electrical resistance and Raman measurements are done on the graphene hole array with P= 600 nm and D= 400 nm.

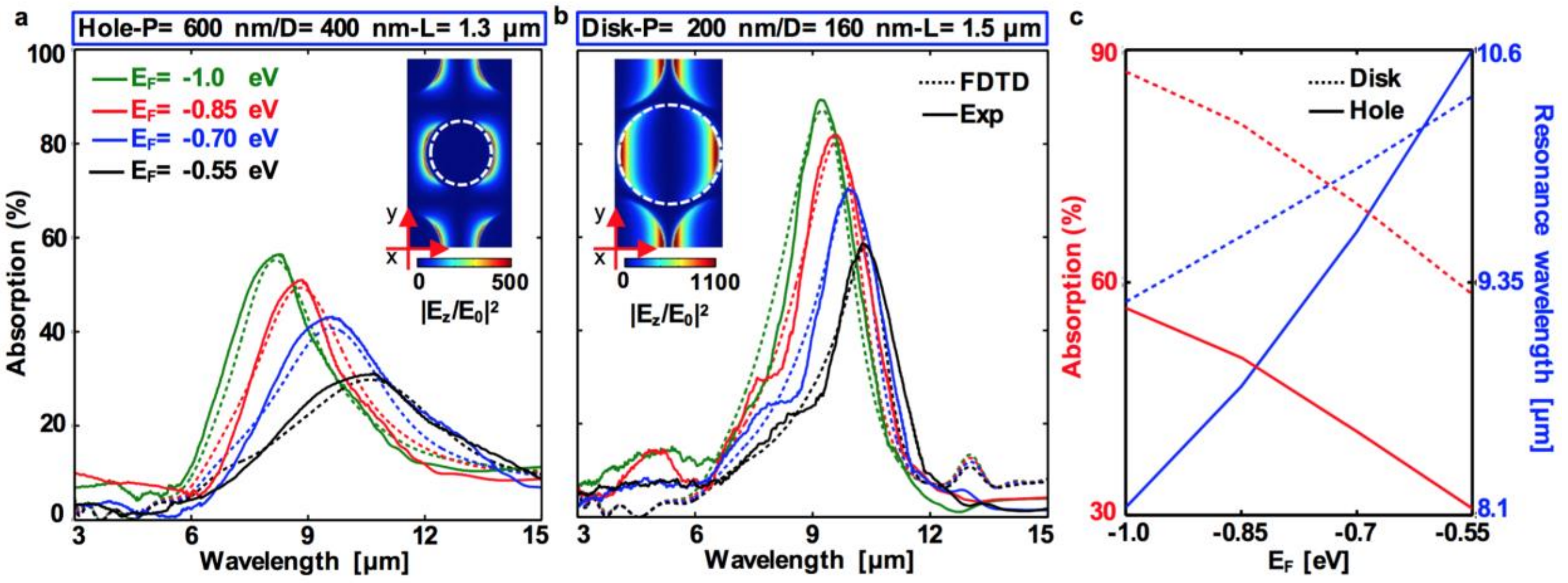

**Figure 4| Experimental results of enhanced light-matter interaction.** (a) The simulated and measured light absorption of the cavity-coupled graphene nanohole array with P=600 nm, D=400 nm and L=1.3 μm for different Fermi energies. (inset) The top view of the z-component near-field profile corresponds to the $E_F$= -1.0 eV. (b) The simulated and measured light absorption of the cavity-coupled graphene nanodisk array with P=200 nm, D=160 nm and L=1.5 μm for different Fermi energies. (inset) The top view of the z-component near-field profile corresponds to the $E_F$= -1.0 eV. (c) The resonance wavelength and corresponding light absorption of the cavity-coupled graphene nanodisk and nanohole arrays as a function of Fermi energy.

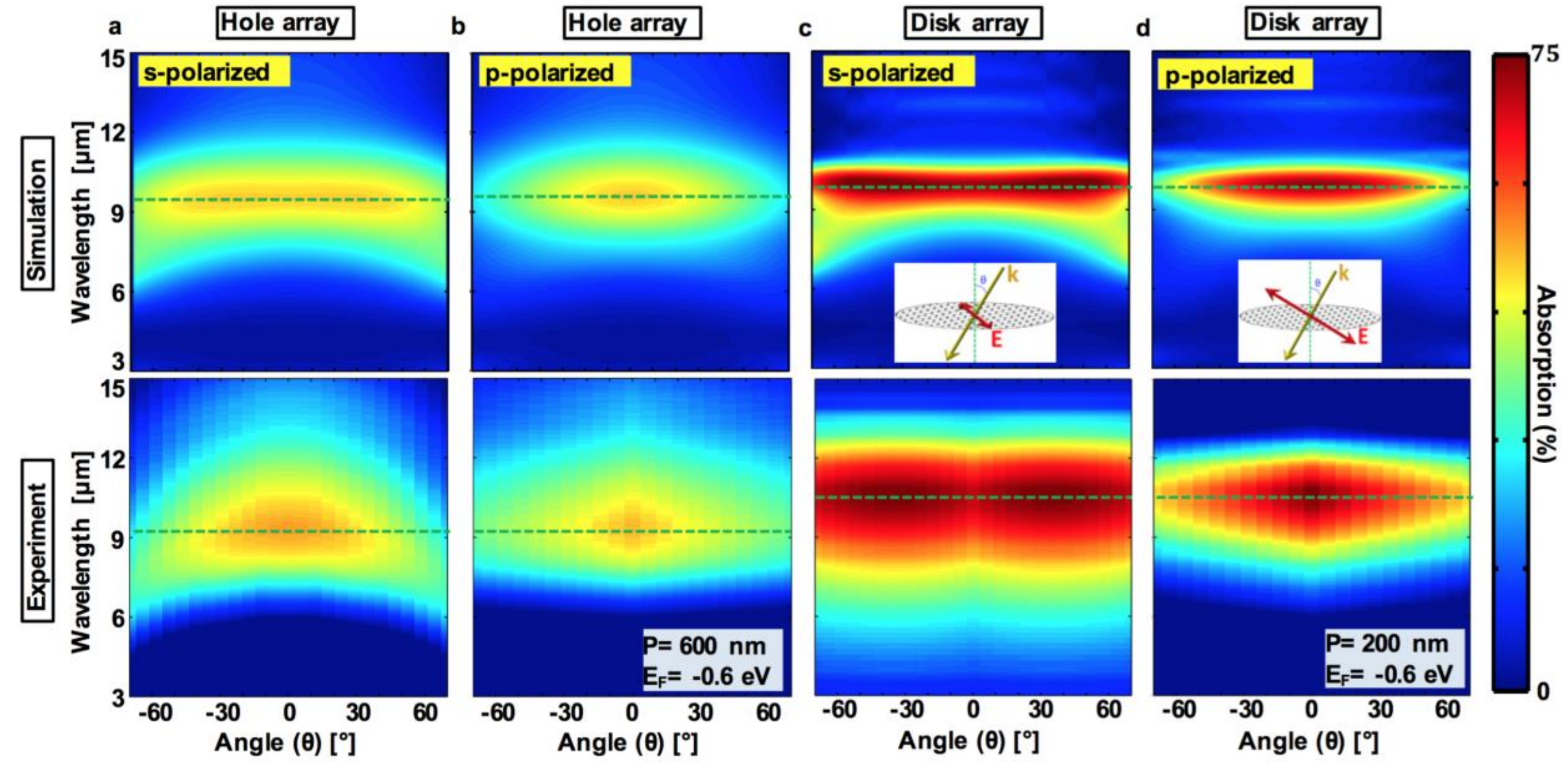

**Figure 5| Angle dependent infrared absorption.** The simulation (top) and experimental (bottom) results for the angular light absorption of the cavity-coupled graphene nanohole (a-b) and nanodisk (c-d) arrays with $E_F$= -0.6 eV as a function of angle of incidence. Inset shows the schematic of the s- and p-polarized incident light. The polarized light measurements are normalized to the light intensity after the polarizer.

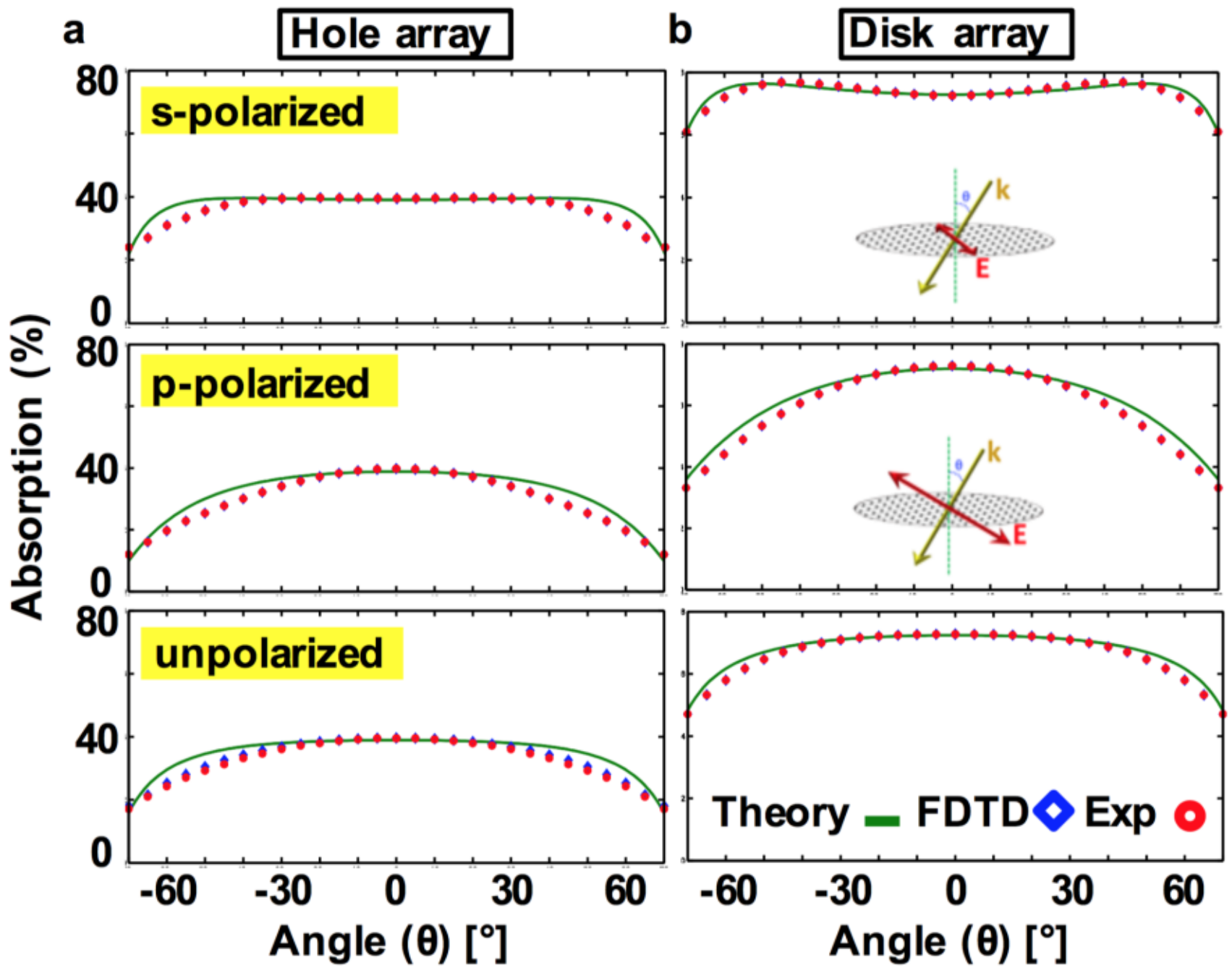

**Figure 6| The angular response for the polarized and unpolarized lights.** Comparison of the analytical modeling, simulation and experimental results of s-polarized (top) and p-polarized (middle) and unpolarized (down) incident light for the graphene nanohole (a) and nanodisk (b) arrays with $E_F$= -0.6 eV. The polarized light measurements are normalized to the light intensity after the polarizer.

Supporting Information for

# Wide Angle Dynamically Tunable Enhanced Infrared Absorption in Large Area Nanopatterned Graphene

*Alireza Safaei[1,2ǂ], Sayan Chandra[2ǂ], Michael N. Leuenberger[1,2,3], Debashis Chanda[1,2,3]**

[1]Department of Physics, University of Central Florida, Orlando, Florida 32816, USA.

[2]NanoScience Technology Center, University of Central Florida, Orlando, Florida 32816, USA.

[3]CREOL, The College of Optics and Photonics, University of Central Florida, Orlando, Florida 32816, USA.

ǂ These authors contributed equally.

*Correspondence and requests for materials should be addressed to D.C. (email: Debashis.Chanda@ucf.edu).

Coupling an optical cavity to patterned graphene, as demonstrated schematically in Fig. S1a, creates constructive/destructive interference of the incident and reflected electric fields on the patterned graphene at quarter ($L = (2m+1)\lambda/4n_{eff}$)/half ($L = m\lambda/2n_{eff}$) wavelength condition to intensify/weaken the localized surface plasmon, as shown in Fig. S1 which is the FDTD-simulation result of the absorption spectra of the cavity-coupled graphene nanohole (b) and nanodisk (c) arrays for different cavity thicknesses at $|E_F| = 1$ eV. The solid white/yellow lines demonstrate the analytically calculated constructive/destructive interference for different cavity modes ($m$) by using the effective refractive index of the cavity spacer in presence of graphene plasmon ($n_{eff}$) calculated by the effective medium theory[1, 2]. The presence of optical cavity doesn't change the localized surface plasmon resonance (LSPR) frequency, as shown in the left panels of Figs. S1b and S1c which are the simulated light absorption spectra of the patterned graphene without optical cavity obtained by using finite-difference time-domain (FDTD) and coupled-dipole approximation[3, 4] (CDA) approaches. Since the particle polarizability used for the nanohole array is that of the complementary nanodisk which has lower loss than nanohole, the CDA predicted absorption is sharper than the FDTD simulation results. The top view of the electric field intensities for the graphene nanodisk and nanohole pattern in Figs. S1d and S1e show the amount of electric field enhancement on the edges which give rise to amplified light-matter interaction.

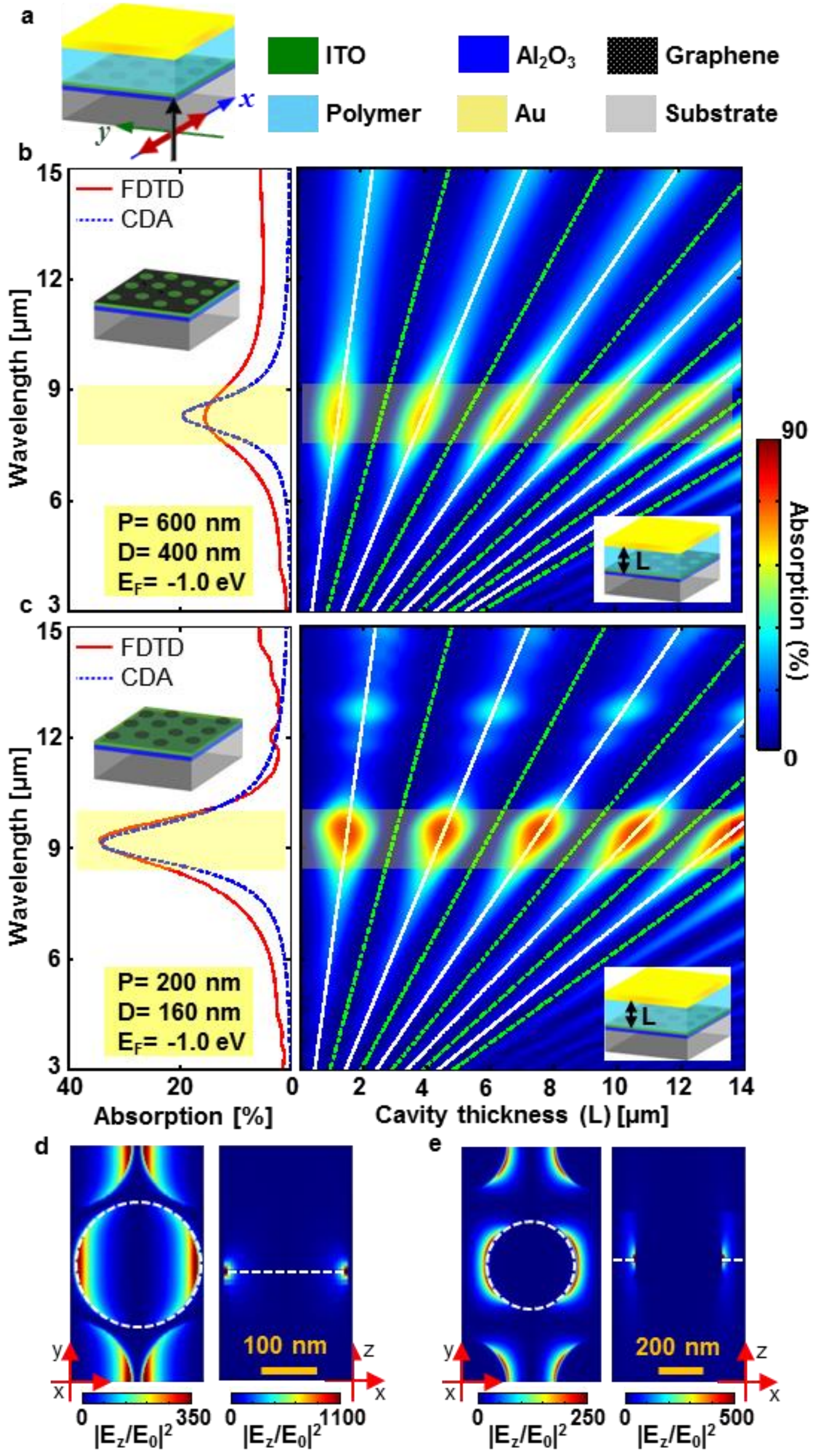

**Figure S1| Exciting surface plasmon on graphene nanodisk and nanohole arrays.** (a) Schematic of the optical cavity-coupled patterned graphene. The light absorption spectra of the optical cavity-coupled graphene nanohole array with P= 600 nm, D= 400 nm, $E_F$= -1 eV (b) and nanodisk array with P= 200 nm, D= 160 nm and $E_F$= -1 eV (c) arrays. The left panels show the simulated light absorption spectra of the nanopatterned graphene arrays obtained by FDTD (solid red) and CDA (dashed blue) approaches. The top view and side view of the z-component electric field intensity of the graphene nanodisk (d) and nanohole (e) arrays.

To investigate more the edge effect on the LSPR, the size of the edge can be changed while the edge-to-edge distance (P-D) remains constant. Decrease in the period P (600 nm/500 nm/400 nm) and diameter D (400 nm/300 nm/200 nm) while their difference (P-D) is constant lead to a blue shift in LSPR frequency and lowers the amount of light absorption in spite of the fact that the trend of the absorption spectra for different cavity thicknesses are similar, as shown in Fig. S2.

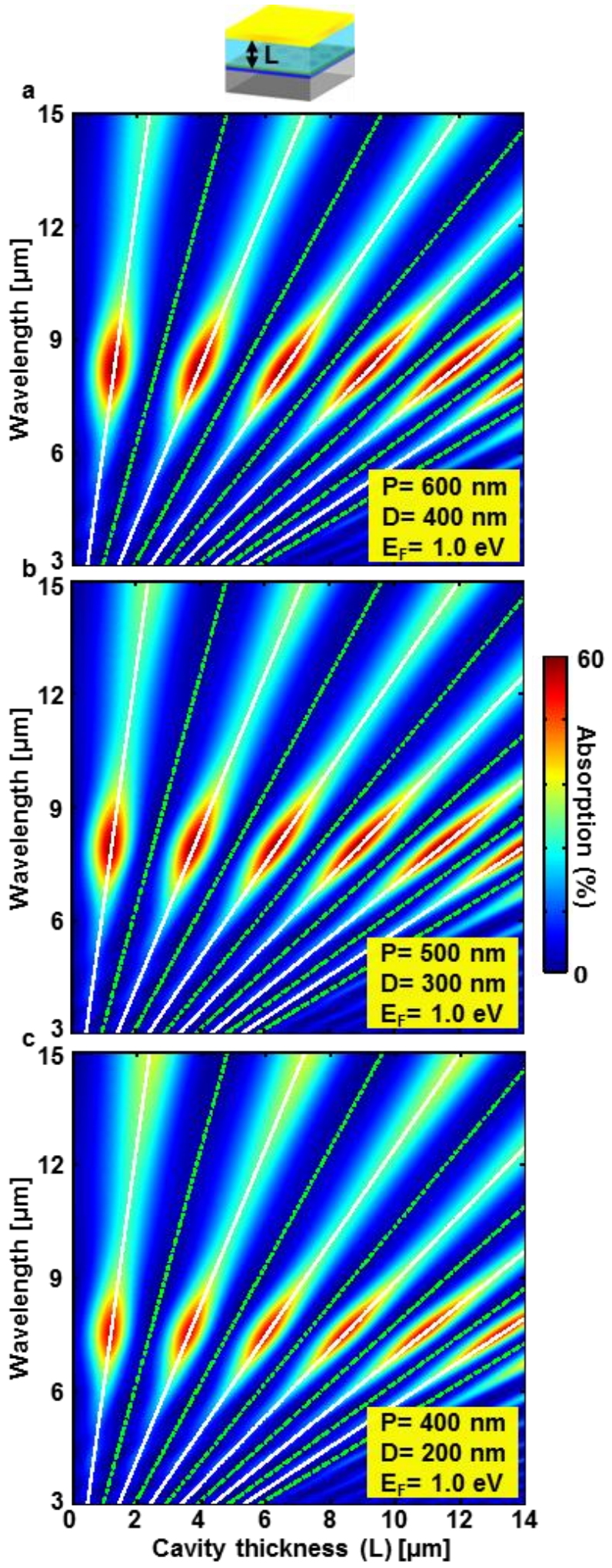

**Figure S2| Exciting surface plasmon on graphene nanohole arrays with different periods and diameters.** The light absorption spectra versus different cavity thicknesses for graphene nanohole array and P= 600 nm, D= 400 nm (a), P= 500 nm, D= 300 nm (b) and P= 400 nm, D= 200 nm (c). The Fermi energy for all simulations is $|E_F|$= 1eV.

Since in the fabricated samples light is incident from the silicon side, the light transmission of silicon wafer with thickness of 100 μm should be high enough. The measured transmission spectrum shows that the average light transmission of the doped silicon substrate in mid infrared wavelength regime is ~ 70%, as shown in Fig. S3.

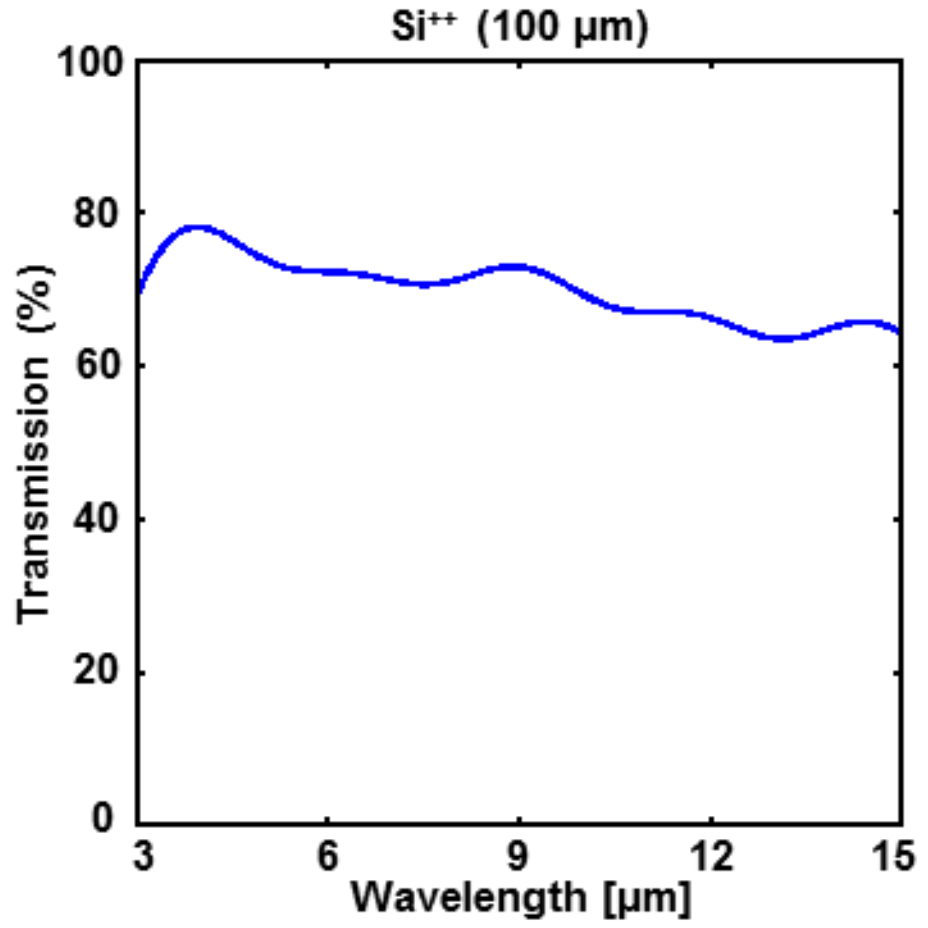

**Figure S3|** The light transmission spectrum of $Si^{++}$ with thickness of 100 μm.

Figure. S4 demonstrates the experimental and the corresponding simulated results for the light absorption spectra of the nanohole (Fig. S4a) and nanodisk (Fig. S4b) patterned graphene without optical cavity which were obtained via the measured reflection (R) and transmission (T) spectra (A= 1-T-R). The light reflection/transmission from the absorber stack without patterned graphene, *i.e.* $Si^{++}$ (100 μm)/$Al_2O_3$ (15 nm)/ITO (30 nm) was taken as the reference for the experimental measurement. The good agreement of simulated and measured spectra validates the experimental results.

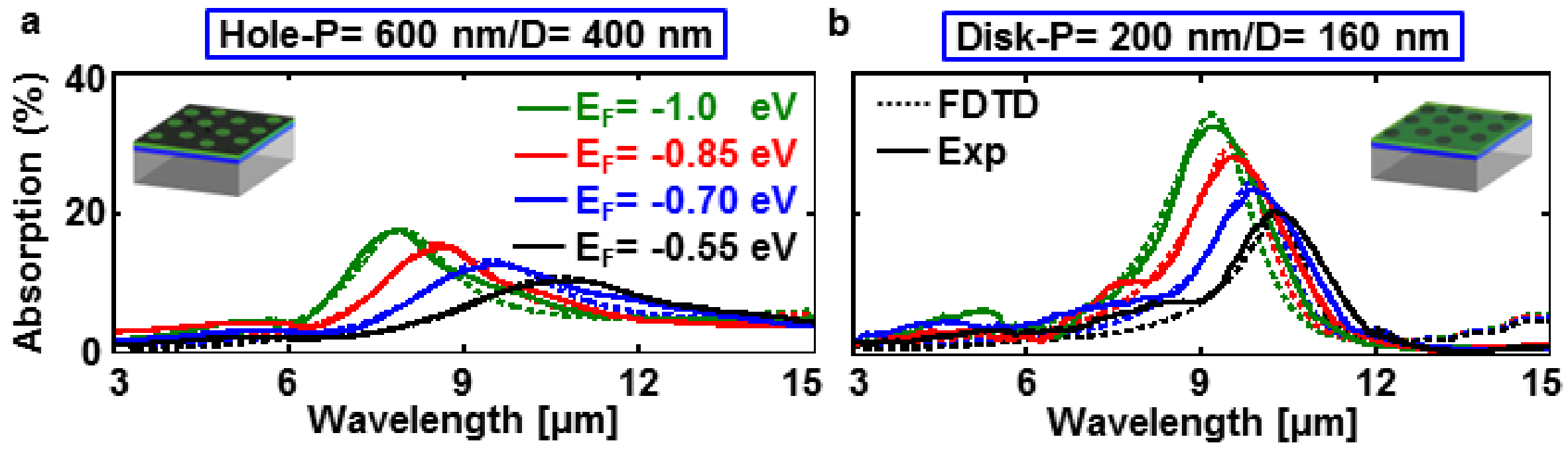

**Figure S4| Experimental results of the enhanced light-matter interaction.** (a) The simulated and measured light absorption of the patterned graphene nanohole array with P=600 nm, D=400 nm for different Fermi energies. (b) The simulated and measured light absorption of the patterned graphene nanodisk array with P=200 nm, D=160 nm for different Fermi energies.

As discussed earlier, reduction in period P and diameter D, while edge-to-edge distance (P-D) is constant give rise to a blue shift in the absorption spectra along with degrading in the absorption, as shown in Figs. S5a for the nanohole graphene array without (top) and with (bottom) optical cavity. The inset shows the trend of LSPR wavelength as period P is decreased. Figure. S5b shows the scanning electron microscope (SEM) images of the fabricated patterned graphene nanohole array.

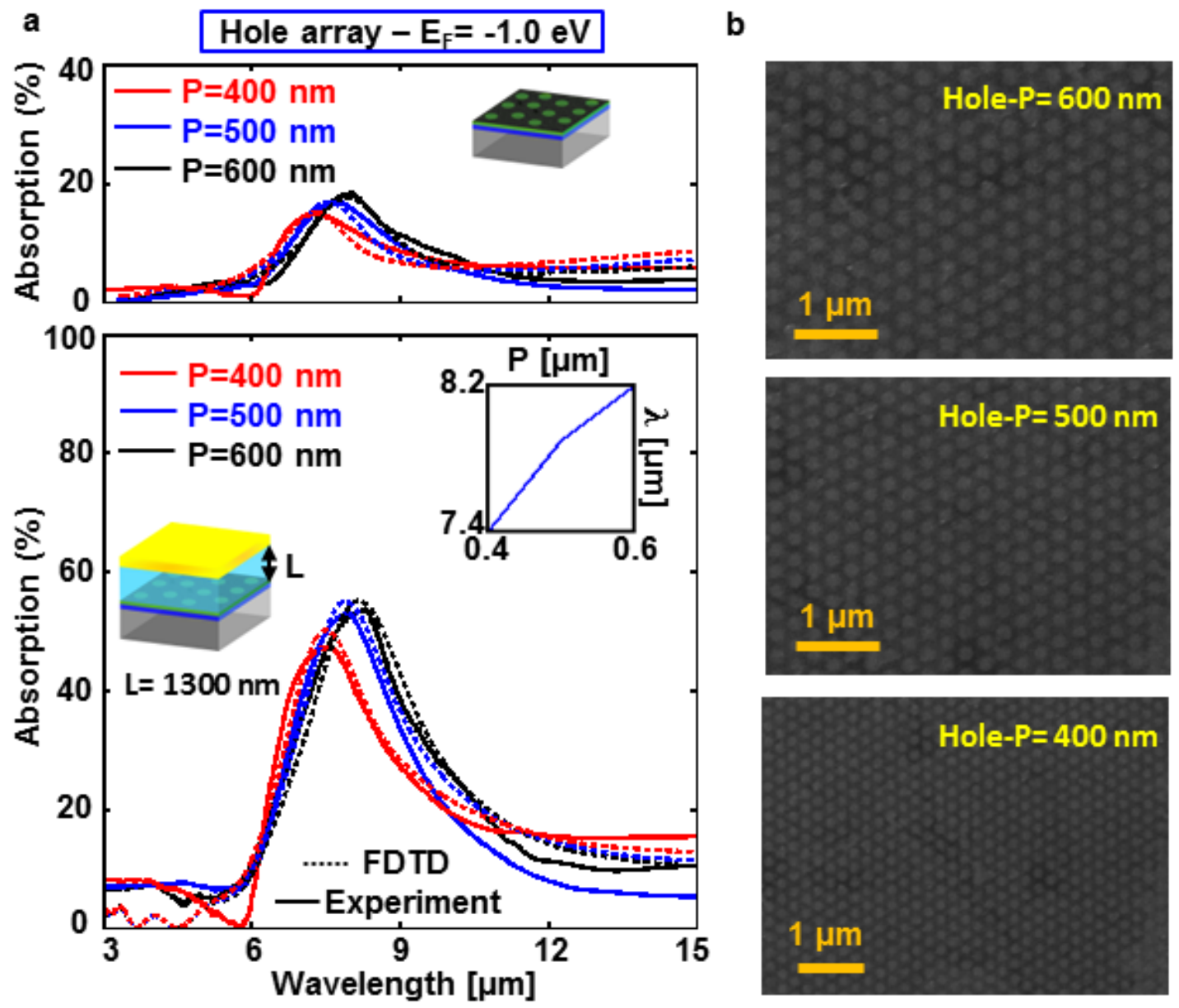

**Figure S5| Exciting surface plasmon on the patterned graphene with nanohole arrays.** (a) The simulated and measured light absorption of the patterned graphene nanohole arrays without (top) and with (bottom) optical cavity with different P and D values. The inset diagram shows the absorption peak wavelength versus period P. The Fermi energy is $E_F$= -1 eV. (b) The SEM images of the fabricated graphene nanohole arrays.

The light absorption spectra of the patterned graphene with and without cavity for different angles of incidence are shown in Figs. S6 (nanohole array-P=600 nm) and S7 (nanodisk array-P=200 nm). The simulated (a-b) and measured (c-d) results show that the general behavior of cavity coupled (a and c) and uncoupled (b and d) patterned graphene are similar for both s-polarized (left panels) and p-polarized (right panels) incident light. The main difference of the samples with and without cavity is related to s-polarized light. As explained in the main manuscript, while the magnitude of electric field parallel to graphene surface remain unaffected for all angles of incidence, the scattering cross-section increases and scales as $\boldsymbol{E}_i sin\theta_i$ and there is an absorption enhancement as $\theta_i$ increases, as shown in Figs S6b,d and S7b,d (left panel). Unlike the no-cavity nanodisk array absorber, as seen from Figs. S6a,c and S7b,c (left panel) the cavity couple system exhibits an increase in absorption for $0^{\circ} < \theta_i < 50^{\circ}$ but for higher angles, because of the interaction between the optical cavity and plasmonic modes resulted in destructive interference of the incident ($\mathbf{E}_i$) and reflected electric fields ($\mathbf{E}_r$) due to phase difference, the absorption drops. The top view of the z-component near-field profiles correspond to $\theta_{inc}= 0^{\circ}$ and $\theta_{inc}= 50^{\circ}$ of the s-polarized light are shown in Fig. S7 which clearly shows different electric dipole magnitudes at those angles.

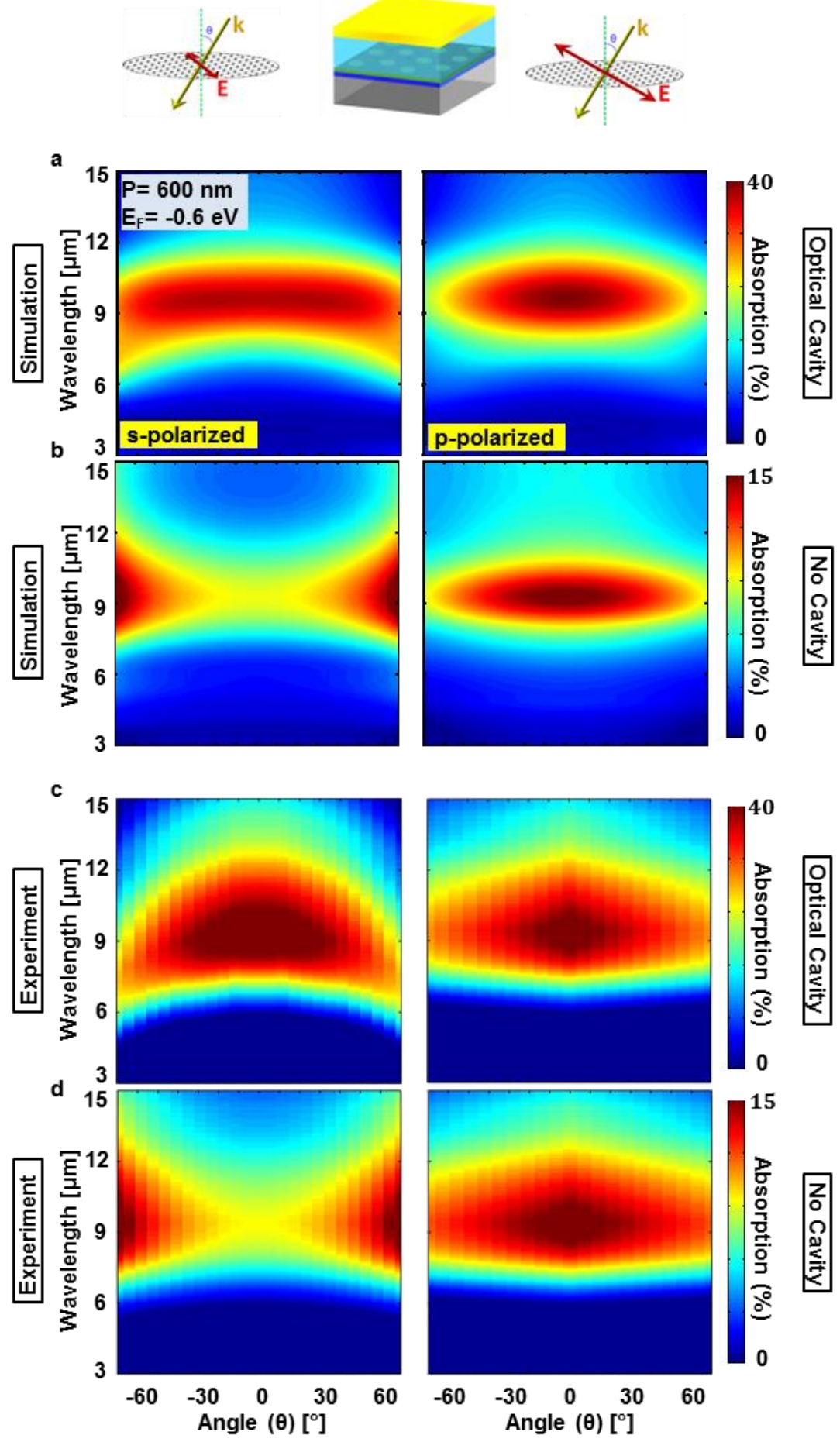

**Figure S6| The absorption of angled incident light by the patterned graphene with nanohole array.** The simulated light absorption of s-polarized (left panel) and p-polarized (right panel) beam by the samples with (a) and without (b) cavity. The corresponding measured light absorption of s-polarized (left panel) and p-polarized (right panel) beam by the samples with (c) and without (d) cavity. The Fermi energy of the graphene samples is $E_F$= -0.6 eV, P= 600 nm and D= 400 nm.

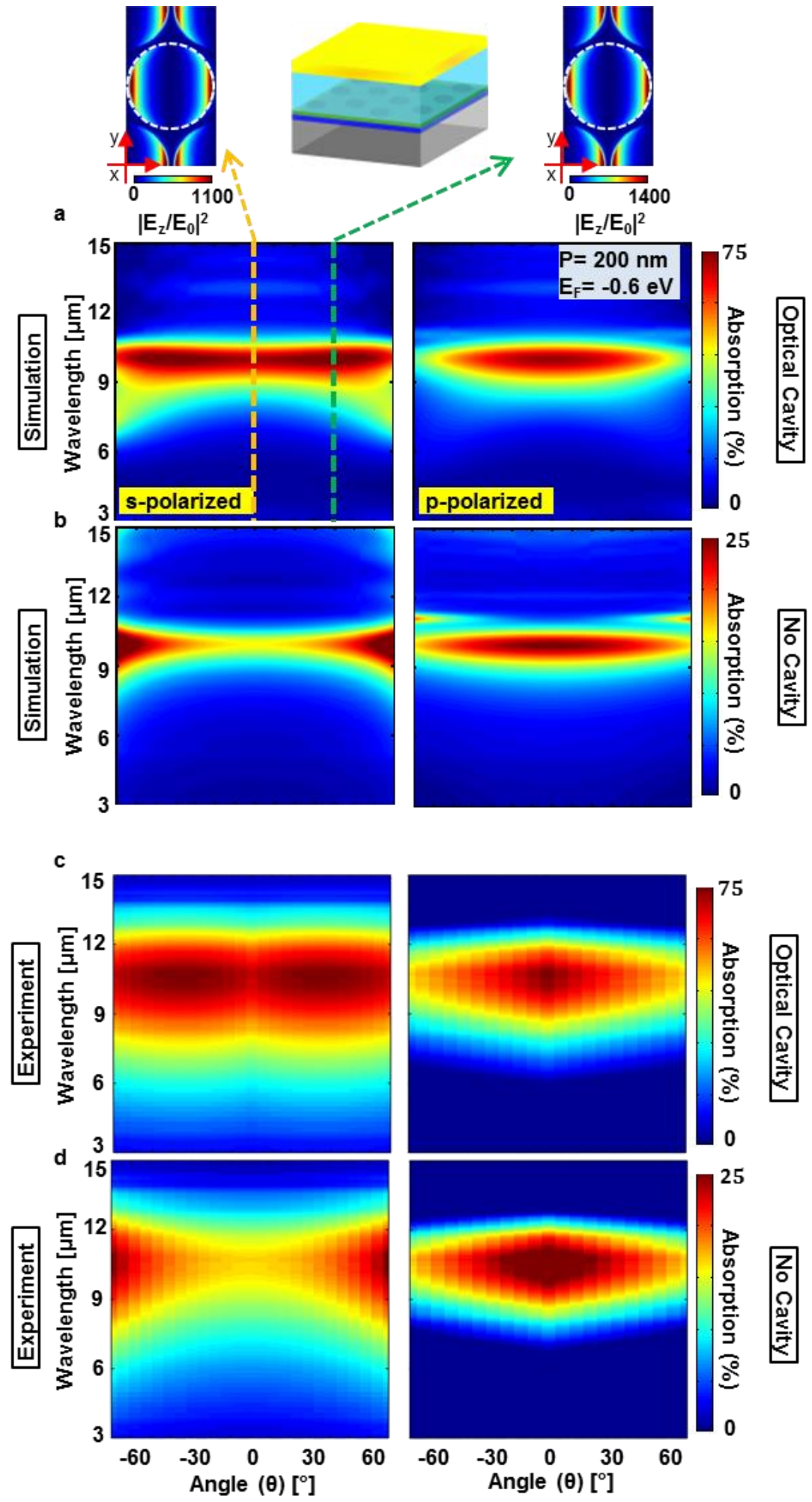

**Figure S7| The absorption of angled incident light by the patterned graphene with nanodisk array.** The simulated light absorption of s-polarized (left panel) and p-polarized (right panel) beam by the samples with (a) and without (b) cavity. The z-component of the nearfield intensity at $\theta_i$= 0$^o$ (left) and $\theta_i$= 50$^o$ (right) of the s-polarized light are shown on top. The corresponding measured light absorption of s-polarized (left panel) and p-polarized (right panel) beam by the samples with (c) and without (d) cavity. The Fermi energy of the graphene samples is $E_F$= -0.6 eV, P= 600 nm and D= 400 nm.

The quality of the high-k gate-dielectric for electrostatic doping of graphene is an important factor for the usability of the designed device. The measured capacitance of the hard-dielectric layer which is 15 nm thick layer of $Al_2O_3$ is $C$ = 0.93 μF/cm$^2$. Such gate dielectric can be used to electrostatically dope the Fermi level of the patterned graphene to $E_F$= -1 eV. The high-k dielectric $Al_2O_3$ has more stability in time compared to the conventionally used soft-ion gel gate dielectric, as the measured light absorption spectra of the corresponding patterned graphene absorbers in a 6-month interval proves this (Figs. S8a-b). The leakage current of the gate-dielectric is another important parameter in power usage which is very low (~$10^{-11}$ A) for 15 nm thick layer of the grown $Al_2O_3$ (Fig. S8c).

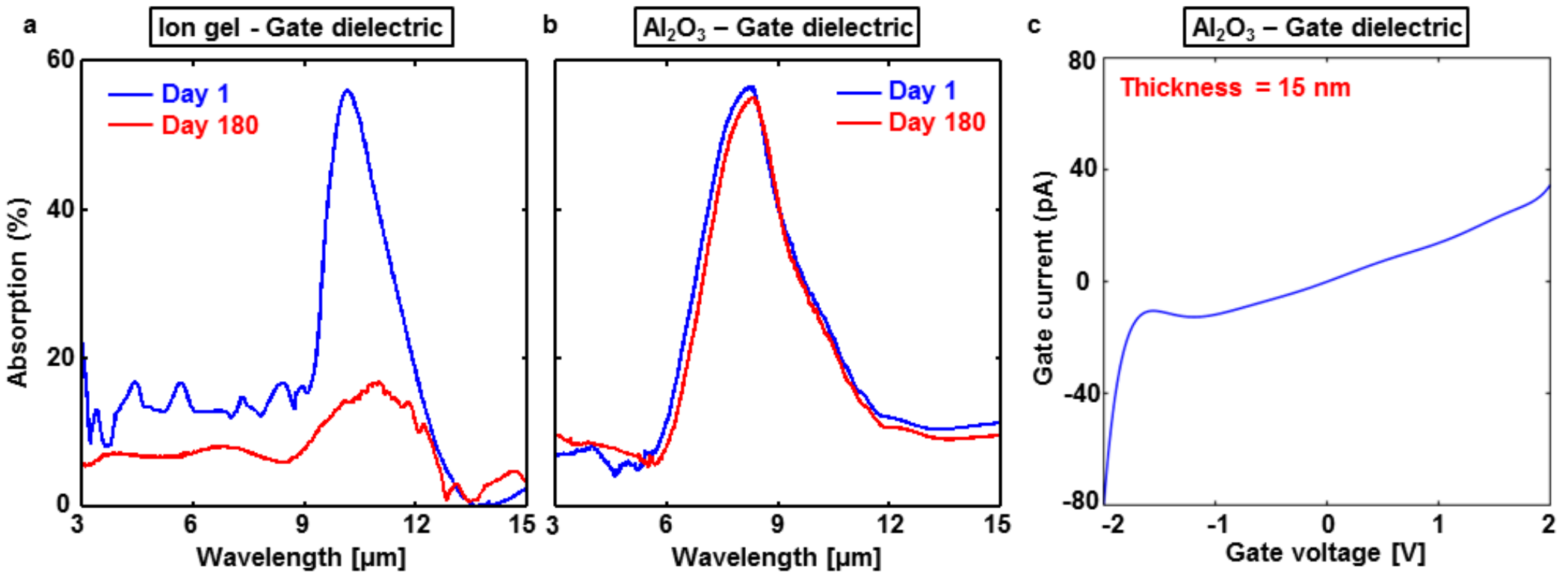

**FigureS8| Characterization and comparison of the hard ($Al_2O_3$) and soft (ion-gel) gate dielectrics.** The light absorption spectra of the fabricated patterned graphene samples doped by ion gel (a) and $Al_2O_3$ (b) in a 6-month time interval. (c) The gate leakage electric current of the 15 nm thick layer of $Al_2O_3$.

To find the experimental value of the carrier mobility $\mu$ of the patterned graphene, the measured electrical resistance $R$ of the patterned graphene is fitted to the theoretical formula ($R = R_0 + 1/\rho e\mu$), where $R_0$ is the minimum resistance at $V_G$= -1 V, $\rho = C\Delta V/e$ is the electron density and $e$ is the Coulomb charge. Based on the diagrams in Fig. S9, the carrier mobility of the patterned graphene is $\mu = 500\ cm^2/V.s$.

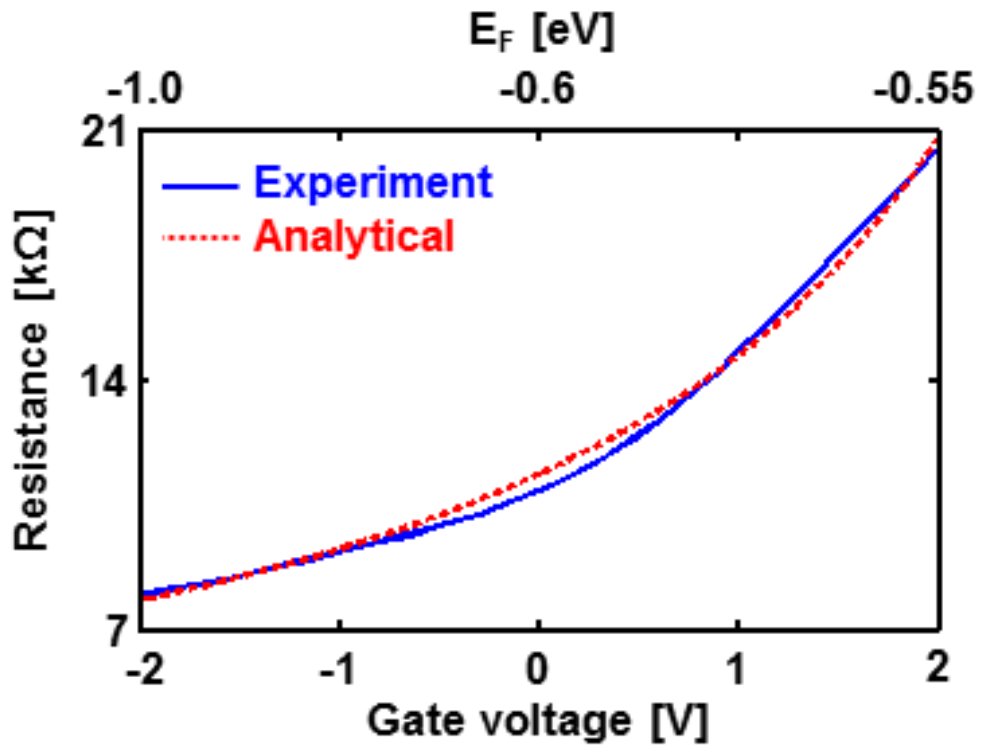

**Figure S9| Electrostatic doping of graphene.** The experimental and analytical model diagrams are overlaid to find the carrier mobility of the patterned graphene sheet.

If the edge-to-edge distance of the patterns in graphene is much smaller than the mean free path (MFP) of electrons and holes, the modified Drude model is not valid to describe the electrical conductivity of graphene [5, 6]. In our case, since the carrier mobility of the transferred graphene is low (μ= 500 $cm^2$/V.s), the edge-to-edge distance for both nanohole and nanodisk arrays are larger than the MFP of electrons and holes ($L_{MFP} = v_F \tau_{DC} = \mu E_F / e v_F < 45$ nm, where $\tau_{DC}$ is the DC momentum relaxation time, $e$ is the electron charge and $v_F$ is the Fermi velocity) and therefore, the Drude model can be applied for the dielectric function calculations and FDTD simulations in this work.

The effect of patterning on Raman spectroscopy of the graphene sheet which is shown in Fig.3 of the main manuscript, can be understood by inserting the boundary condition to the atomic displacement equation $\boldsymbol{u}$, the longitudinal (LO) and transverse (TO) optical phonon dispersions are adjusted, i.e. $\omega_n^2 = \omega_{LO}^2 - \lambda^2(q_n^2 + q^2)^2 + \beta_L^2(q_n^2 + q^2)$ and $\omega_n^2 = \omega_{TO}^2 - \beta_T^2(q_n^2 + q^2)$, where $q_n = n\pi/L$ and $L$ is edge-to-edge distance, λ, $\beta_L$ and $\beta_T$ are the parameters approximated by LO and TO phonon dispersion curves [7, 8]. The displacement equation is written as $\ddot{\boldsymbol{U}} = \hat{\Lambda}^{opt}\boldsymbol{U}$, where $\boldsymbol{U}$ is the relative displacement of two sublattices and $\hat{\Lambda}^{opt}$ is the operator for the graphene optical phonons[7]. The calculations based on the above equations suggest a red shift consistent with that observed experimentally, as shown in Fig. 3c of the main manuscript.

The decay rates which were used in the graphene polarizability $\alpha$ in Eq. 2 were extracted from the simulated extinction cross section per area for a single graphene nanodisk or nanohole and fitting the theoretical extinction cross-section $\sigma^{ext} = (4\pi\omega/c)Im\{\alpha(\omega)\}$ [9] to those curves, as shown in Fig. S10. According to these figures, $\hbar\kappa = 3.5 \times 10^{-2}\ eV/6.9 \times 10^{-2}\ eV$ and $\hbar\kappa_r = 1.22 \times 10^{-4}\ eV/4.5 \times 10^{-4}\ eV$ for graphene nanodisk/nanohole. The plasmon decay rates show that the plasmon lifetime of the graphene nanodisk is higher than graphene nanohole.

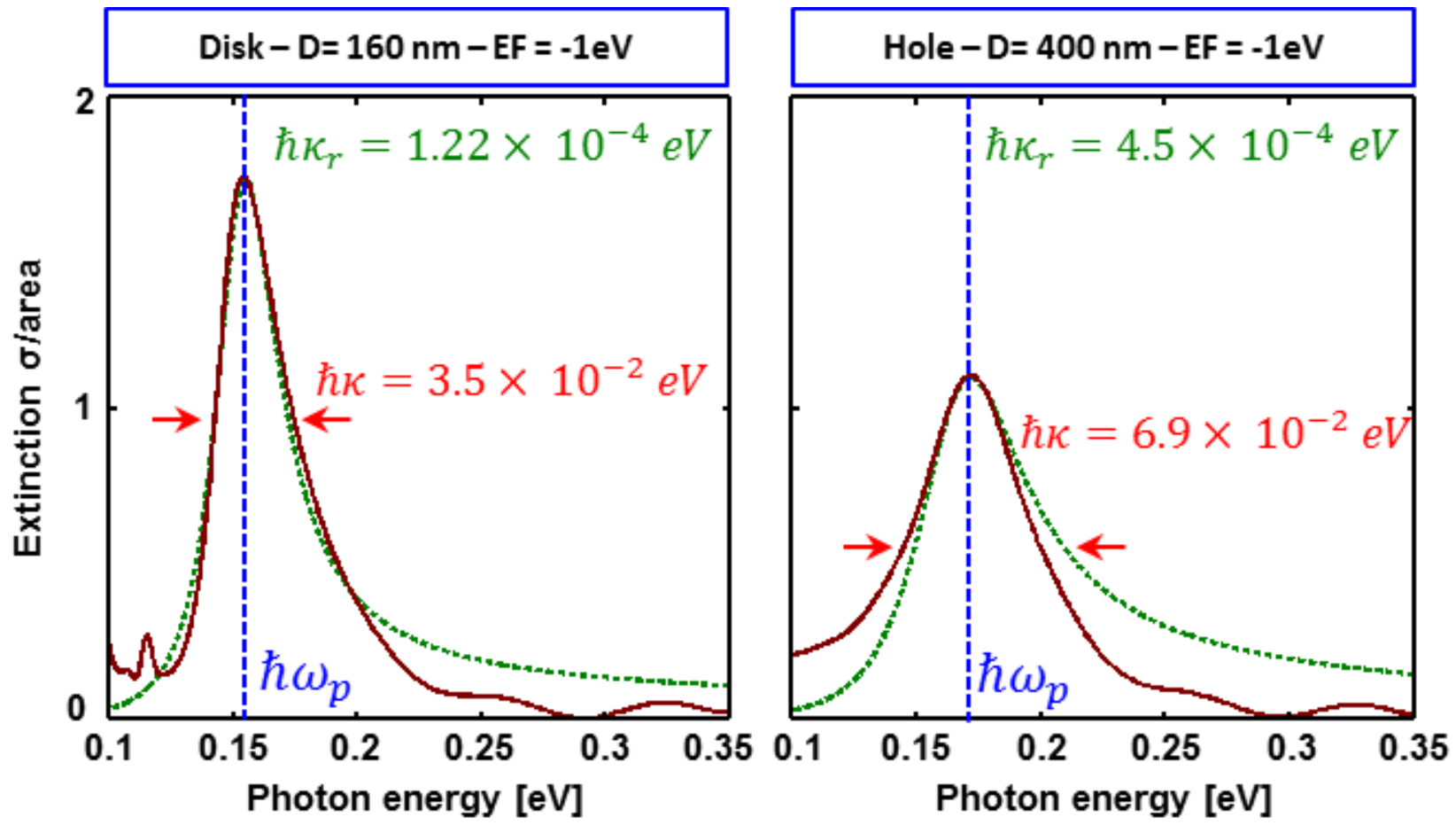

**Figure S10| The plasmon decay rates.** The simulated extinction cross sections per area of a single graphene nanodisk with D= 160 nm (a) and nanohole with D= 400 nm at $E_F$= -1 eV were used to extract the radiative ($\hbar\kappa_r$) and total ($\hbar\kappa$) plasmon decay rates.

## References:

1. Granqvist, C. G.; Hunderi, O. *Physical Review B* **1978,** 18, (4), 1554-1561.
2. Safaei, A.; Chandra, S.; Vázquez-Guardado, A.; Calderon, J.; Franklin, D.; Tetard, L.; Zhai, L.; Leuenberger, M. N.; Chanda, D. *Physical Review B* **2017,** 96, (16).
3. García de Abajo, F. J. *Reviews of Modern Physics* **2007,** 79, (4), 1267-1290.
4. Maier, S. A. **2007**.
5. Falkovsky, L. A.; Pershoguba, S. S. *Physical Review B* **2007,** 76, (15).
6. Gao, W.; Shu, J.; Qiu, C.; Xu, Q. *ACS nano* **2012,** 6, (9), 7806-13.
7. Qian, J.; Allen, M. J.; Yang, Y.; Dutta, M.; Stroscio, M. A. *Superlattices and Microstructures* **2009,** 46, (6), 881-888.
8. Maultzsch, J.; Reich, S.; Thomsen, C.; Requardt, H.; Ordejon, P. *Phys Rev Lett* **2004,** 92, (7), 075501.
9. Thongrattanasiri, S.; Koppens, F. H. L.; García de Abajo, F. J. *Physical Review Letters* **2012,** 108, (4).